\documentclass[twocolumn,floatfix,pra,aps,showpacs,superscriptaddress]{revtex4}
\usepackage{epsfig,graphicx,tabularx}
\usepackage{amsmath}
\usepackage{color,braket}

\newcommand{\mb}{\mathbf}
\newcommand{\be}{\begin{equation}}
\newcommand{\ee}{\end{equation}}

\begin{document}

\title{Persistent currents in two-component condensates in a toroidal trap}

\author{M. Abad}
\affiliation{INO-CNR BEC Center and Dipartimento di Fisica, Universit\`a di Trento, 38123 Povo, Italy}
\author{A. Sartori}
\affiliation{INO-CNR BEC Center and Dipartimento di Fisica, Universit\`a di Trento, 38123 Povo, Italy}
\author{S. Finazzi}
\affiliation{INO-CNR BEC Center and Dipartimento di Fisica, Universit\`a di Trento, 38123 Povo, Italy}
\affiliation{Laboratoire Mat\'eriaux et Ph\'enom\`enes Quantiques, Universit\'e Paris Diderot-Paris 7 and CNRS, B\^atiment Condorcet, 75205 Paris Cedex 13, France
}
\author{A. Recati}
\affiliation{INO-CNR BEC Center and Dipartimento di Fisica, Universit\`a di Trento, 38123 Povo, Italy}

\begin{abstract}
The stability of persistent currents in a two-component Bose-Einstein condensate in a toroidal trap is studied both in the miscible and immiscible regimes. 
In the miscible regime we show that superflow decay is related to linear instabilities of the spin-density Bogoliubov mode. We find a region of partial stability, where the flow is stable in the majority component while it decays in the minority component. We also characterize the dynamical instability appearing for a large relative velocity between the two components. In the immiscible regime the stability criterion is modified and depends on the specific density distribution of the two components. The effect of a coherent coupling between the two components is also discussed.
\end{abstract}


\maketitle

\section{Introduction}
Persistent currents are dissipationless flows representing one of the strongest signatures of superfluidity. They are topological long-lived metastable states of quantum fluids which are described by a macroscopic wave function (order parameter in Bose-Einstein condensates).
Persistent currents become unstable above a certain velocity threshold \cite{Bloch}.
In the absence of a weak-link, their decay is a complex, stochastic process mediated by phase slips \cite{Moulder2012}, and is related to the existence of energy barriers for the excitations to cross the bulk superfluid \cite{Benakli,Tempere,Dora}. 

The versatility of gaseous Bose-Einstein condensates (BECs) and the experimentalists' ability to control their properties offer new scenarios to probe superfluidity. One of the most intriguing systems that can be realized nowadays is the spinor condensate, which is described by a vectorial order parameter.
The simplest example is the two-component condensate, where spin exchange can be implemented by a coherent coupling between internal levels of the atoms. This system has recently acquired new relevance as the basis for BECs with spin-orbit coupling.

In the absence of coherent coupling, a two-component condensate is usually referred to as a binary mixture.
Such a system shows two possible ground states with different symmetries, depending on whether the mixture is miscible (homogeneous phase) or immiscible (phase separated). Whereas the features of the phase transition have been deeply studied both theoretically and experimentally, the superfluid properties of the mixture are still controversial, especially regarding the stability of persistent currents. Indeed existing theoretical predictions~\cite{Smyrnakis2009,Bargi2010,Anoshkin2013,Wu2013,Smyrnakis2014} do not explain the recent experimental observations in \cite{Beattie2013}. Also, very recently the dynamics of the persistent currents have been numerically simulated using spin-1 Gross-Pitaevskii equations~\cite{Yakimenko}, but a deep theoretical understanding of the results in \cite{Beattie2013} is  missing.
Furthermore, arguments related to the continuous twisting of the order parameter \cite{Leggett} can be neither applied to the mixture configuration nor in the presence of coherent coupling because in both cases the Hamiltonian is generally not invariant under SU(2) transformations.

In this work we study the microscopic mechanism that triggers the decay of persistent currents and we build the stability diagram in a quasi-two-dimensional (2D) ring geometry. In the miscible regime our theoretical analysis is based on the solution of Bogoliubov excitations and it is addressed numerically both with imaginary-time and real-time simulations. We show that there exists a regime of partial instability where the minority component could lose angular momentum without affecting the majority component. The existence of this regime is the main result of this work and could be at the origin of the experimental observations in \cite{Beattie2013}, although to test it fully new experiments should be carried out with different parameters. We also discuss the stability conditions in the phase separated regime and in the presence of a coherent coupling between the two components.
%Our reasoning it is also in agreement with very recent results and the recent numerical results of \cite{Yakimenko}, although more experiments should be carried out to fully test our predictions.

The article is organized as follows. In Sec.~\ref{SecTheory} we describe the system under consideration and we settle the theoretical framework. In Sec.~\ref{SecBog} we derive the dispersion relations of Bogoliubov excitations for a binary mixture, which are at the basis of the stability criterion. In Sec.~\ref{SecCorrSound} we calculate the main correction to the sound velocity due to confinement. Section~\ref{SecMisc} is devoted to the stability of persistent currents in the miscible regime of the mixture. We present the stability diagram of persistent currents predicted by imaginary-time simulations of the Gross-Pitaevskii equations in Sec.~\ref{SecNum}. The physical origin of the partially stable region is discussed in~\ref{SecPartStab} using a linear stability analisis, which we confirm with real-time dynamics simulations. In Sec.~\ref{SecDynInst} we characterize the dynamical instability known as counterflow instability. The stability of persistent currents in the phase-separated regime is analyzed in Sec.~\ref{SecPhSep}, and the effect of adding a coherent coupling is discussed in Sec.~\ref{SecCoherent}. Finally, the conclusions are drawn in Sec.~\ref{SecConclusions}.

\section{System description and theoretical framework}\label{SecTheory}

We consider a two-component condensate strongly confined along the longitudinal direction, $z$, such that the dynamics is effectively two-dimensional. For concreteness we assume a harmonic confinement in this direction, $V_z=m\omega_z^2 z^2/2$, with $\omega_z$ the trapping frequency and $m$ the atomic mass. In the 2D limit we are considering, $\hbar\omega_z$ must be much larger that all the other energy scales. At the mean field level, this system is described by two wavefunctions (order parameters) $\Psi_a$ and $\Psi_b$, 
normalized to the number of particles in each species, respectively $N_a$ and $N_b$.
The wave functions satisfy the coupled Gross-Pitaevskii (GP) equations
\begin{align}
  i\hbar \frac{\partial }{\partial t}\Psi_a = &\left[-\frac{\hbar^2}{2m}\nabla_\perp^2 + V + g_{a} |\Psi_a|^2 + g_{ab}|\Psi_b|^2 \right]\Psi_a \label{tdgpa},\\
  i\hbar \frac{\partial }{\partial t}\Psi_b = &\left[-\frac{\hbar^2}{2m}\nabla_\perp^2 + V + g_{b} |\Psi_b|^2 + g_{ab}|\Psi_a|^2 \right]\Psi_b \label{tdgpb},
\end{align} 
where $V$ is a ring-shaped external potential obtained as the sum of harmonic and Gaussian potentials,
\be
 V=\frac{1}{2}m\omega_\perp^2r_\perp^2+ V_0e^{-2r_\perp^2/\sigma_0^2}\ ,\label{EqPot}
\ee
with $\omega_\perp$ the radial trapping frequency,  $r_\perp^2=x^2+y^2$ the radial coordinate, $\sigma_0$ the beam waist, and $V_0$ the strength of the laser beam, which is proportional to its intensity. This choice for the potential follows the experiments \cite{Ramanathan2011,Wright2013,Murray2013}. A different choice -- as for instance in \cite{Moulder2012,Beattie2013} -- might slightly change our results quantitatively, but not qualitatively. The interatomic interactions are characterized by the intra- ($g_{a}$, $g_{b}$) and inter-species ($g_{ab}$) coupling constants. They are given in terms of the three dimensional (3D) $s$-wave scattering lengths, $a^{3D}$, through $g/(\hbar\omega_\perp a_\perp^2)=\sqrt{8\pi\lambda}\,a^{3D}/a_\perp$, where $\lambda=\omega_z/\omega_\perp$ is the trap aspect ratio, and $a_\perp=\sqrt{\hbar/m\omega_\perp}$ is the transverse harmonic oscillator length. 

Since particle exchange is forbidden in Eqs.~(\ref{tdgpa})-(\ref{tdgpb}) the number of particles in components $a$ and $b$ is fixed externally. It is then convenient to introduce the polarization of the mixture, $P_z=(N_a-N_b)/N$, with $N=N_a+N_b$ the total number of atoms. Different polarizations are achieved experimentally by coherently coupling the two components for a controlled time (see for instance Ref.~\cite{Beattie2013}). Throughout this work we will analyze the stability properties in terms of the polarization $P_z$.

The above GP equations admit stationary solutions $\Psi_\sigma(\mathbf{r},t)=e^{-i\mu_\sigma t/\hbar}\psi_\sigma(\mathbf{r})$, with $\sigma=a,b$. The chemical potentials $\mu_a$ and $\mu_b$ generally differ because $N_a$ and $N_b$ are conserved separately.
A first-order phase transition at $g_{ab}^c=\sqrt{g_ag_b}$ separates two possible ground states: for $g_{ab}<g_{ab}^c$ the mixture is miscible and both gases occupy the same volume, whereas when $g_{ab}>g_{ab}^c$ the two components $a$ and $b$ separate into spatial domains (phase-separated or immiscible regime). In the latter case, the density distribution depends on the geometry of $V$, as well as the mass ratio and the interaction ratio $g_{a}/g_{b}$ (see for instance~\cite{Svidzinsky}). In Bose gases like $^{87}$Rb, with nearly equal intra- and inter-species interaction strengths, the system is usually on the verge of instability. 
%Two possible ground states separated by a first-order phase transition are possible. Indeed there exists a critical value $g_{ab}^c=\sqrt{g_ag_b}$ of the interspecies coupling constant, such that, when $g_{ab}<g_{ab}^c$, the mixture is miscible and the ground state is homogeneous, whereas when $g_{ab}>g_{ab}^c$ the two components $a$ and $b$ separate into spatial domains (phase-separated or immiscible regime). As a consequence, Bose gases like $^{87}$Rb with nearly equal intra- and inter-species interaction strengths are usually on the verge of instability. 

Unless otherwise stated we consider the following parameters in the numerical simulations: $N=10^5$, $V_0/\hbar\omega_\perp=220$, $\sigma_0/a_\perp=6$, $g_a=g_b=g$, $g_{ab}/g=0.97$ ($g_{ab}/g=1.02$ in phase separation). These values of $g$ would correspond in the case of $^{87}$Rb to $a^{3D}_a=a^{3D}_b\approx100.0\,a_B$ and $a^{3D}_{ab}\approx97.3\,a_B$ ($a^{3D}_{ab}\approx101.7\,a_B$ in phase separation), where $a_B$ is the Bohr radius, for a trapping potential with $\lambda\approx88$, with $\omega_\perp=2\pi\times50$~Hz.
All simulations have been carried out in a mesh of $256\times256$ points with a grid spacing $h_{x}= h_{y}\approx 0.2~a_{\perp}$.
The algorithm we have used for real-time propagation is based on the split-operator method, and the kinetic term is treated in Fourier space. We have also run simulations based on a Hamming's algorithm (predictor-corrector-modifier) initialized by a 4th-order Runge Kutta, and we have found very good agreement between the two methods.

\subsection{Bogoliubov excitations in uniform medium}\label{SecBog}

Since there exist two regimes (miscible and phase separated) with different spatial properties, their excitation spectra will be different, and also the corresponding stability criteria. Indirectly, this has been seen in the theoretical analysis of \cite{Shimodaira2010}, where the miscible-immiscible transition was scanned in the presence of rotation.

In the ground state of the phase separated regime, there are two single-species condensates separated by a domain wall. Excluding the modes localized in the vicinity of the wall~\cite{Mazets2002,Takeuchi2013}, the Bogoliubov excitation spectrum in the bulk is that of the single component. As we will discuss in Sec.~\ref{SecPhSep} the stability of persistent currents in the immiscible regime is related to the appearance of a barrier created by the minority component.

In the miscible regime, in contrast, the stability of persistent currents is closely related to the Bogoliubov spectrum. In this section we calculate it for a uniform ($V=0$) system and the effect of the transverse degrees of freedom will be incorporated in Sec.~\ref{SecCorrSound}. %The spectrum of a binary mixture with the two components $a$ and $b$ moving at velocities $\mathbf{v_a}$ and $\mathbf{v_b}$, respectively, can be calculated using Bogoliubov prescription
Since we are interested in the stability of states with superfluid currents we need to calculate the Bogoliubov spectrum above a mean-field state where components $a$ and $b$ move at velocities $\mathbf{v_a}$ and $\mathbf{v_b}$, respectively, 
\begin{equation}
   \Psi_{\sigma}(\mathbf{r},t)=\left(\psi_{\sigma}+\delta\Psi_{\sigma}(\mathbf{r},t)\right)e^{-i\mu_{\sigma} t/\hbar}e^{-im\mathbf{v}_{\sigma}\cdot\mathbf{r}/\hbar}\label{EqBogAnsatz}
\end{equation}   
where $\mu_{a}=mv_a^2/2+g_{a}n_{a}+g_{ab}n_{b}$ is the chemical potential for component $a$ ($\mu_b$ is the corresponding expression for component $b$). 
The superfluid velocities $\mathbf{v_{\sigma}}$ are related to the phase of the order parameters, $S_{\sigma}$, as $\mb{v_{\sigma}}=\hbar/m \nabla S $, where $\sigma=a,b$.
To find the equations for the (small) perturbations $\delta\Psi_\sigma$, we substitute Eq.~(\ref{EqBogAnsatz}) into the GP Eqs.~(\ref{tdgpa}) and (\ref{tdgpb}) and linearize them. The perturbations can be decomposed in a plane-wave basis as 
\begin{align}
&\delta\Psi_{\sigma}\sim \mathcal{U}_{\sigma} e^{i(\mathbf{k}\cdot\mathbf{r}-\omega t)}\\
&\delta\Psi_{\sigma}^*\sim \mathcal{V}_{\sigma} e^{i(\mathbf{k}\cdot\mathbf{r}-\omega t)} 
\end{align}
After some algebra an eigenvalue equation is reached,
%
%\begin{widetext}
%\begin{displaymath}
% \hbar\omega \left(
% \begin{array}{c}
%  u_a \\ v_a\\ u_b \\v_b
% \end{array}\right)=\left( 
% \begin{array}{cccc}
%   h_{a} -\hbar \mathbf{v}_a\cdot\mathbf{k}  & g_{a}n_a &   g_{ab}\sqrt{n_an_b}  & g_{ab}\sqrt{n_an_b} \\
%  -g_{a}n_a & -h_{a} -\hbar \mathbf{v}_a\cdot\mathbf{k}  & -g_{ab}\sqrt{n_an_b} & - g_{ab}\sqrt{n_an_b} \\
%   g_{ab}\sqrt{n_an_b}  & g_{ab}\sqrt{n_an_b} & h_{b} -\hbar \mathbf{v}_b\cdot\mathbf{k} & g_{b}n_b \\
%  -g_{ab}\sqrt{n_an_b} & -g_{ab}\sqrt{n_an_b} & -g_{b}n_b & -h_{b} -\hbar \mathbf{v}_b\cdot\mathbf{k} 
% \end{array}\right)
% \left(
% \begin{array}{c}
%  u_a \\ v_a\\ u_b \\v_b
% \end{array}\right)
%\end{displaymath}
%\end{widetext}
\begin{equation}
 \hbar\omega \left(
 \begin{array}{c}
  \mathcal{U}_a \\ \mathcal{V}_a\\ \mathcal{U}_b \\\mathcal{V}_b
 \end{array}\right)=\mathcal{L}
 \left(
 \begin{array}{c}
  \mathcal{U}_a \\ \mathcal{V}_a\\ \mathcal{U}_b \\\mathcal{V}_b
 \end{array}\right)\ ,\label{EqBog}
\end{equation}
where the linear operator $\mathcal{L}$ is given by
\begin{widetext}
\begin{equation}
 \mathcal{L}=\left( 
 \begin{array}{cccc}
   h_{a} -\hbar \mathbf{v}_a\cdot\mathbf{k}  & g_{a}n_a &   g_{ab}\sqrt{n_an_b}  & g_{ab}\sqrt{n_an_b} \\
  -g_{a}n_a & -h_{a} -\hbar \mathbf{v}_a\cdot\mathbf{k}  & -g_{ab}\sqrt{n_an_b} & - g_{ab}\sqrt{n_an_b} \\
   g_{ab}\sqrt{n_an_b}  & g_{ab}\sqrt{n_an_b} & h_{b} -\hbar \mathbf{v}_b\cdot\mathbf{k} & g_{b}n_b \\
  -g_{ab}\sqrt{n_an_b} & -g_{ab}\sqrt{n_an_b} & -g_{b}n_b & -h_{b} -\hbar \mathbf{v}_b\cdot\mathbf{k} 
 \end{array}\right)\label{EqL}
\end{equation}
\end{widetext}
and where we have defined $h_{\sigma}=\hbar^2 k^2/(2m) + g_{\sigma}n_\sigma$, with $\sigma=a,b$. Diagonalization of $\mathcal{L}$ gives four eigenvalues, and four corresponding eigenvectors. 
Notice that since the linear operator is not hermitician the frequencies might be complex (indeed, when they become complex they give rise to a dynamical instability, which will be further discussed in Sec.~\ref{SecDynInst}). In general two of the eigenvalues have a positive norm, defined as $|\mathcal{U}_{a}|^2-|\mathcal{V}_{a}|^2+|\mathcal{U}_{b}|^2-|\mathcal{V}_{b}|^2$, while the other two have a negative norm. The relative sign of the amplitudes $\mathcal{U}_{a}$ and $\mathcal{U}_{b}$ (and correspondingly $\mathcal{V}_{a}$ and $\mathcal{V}_{b}$), determines whether the modes are in phase (density mode) or out of phase (spin-density mode). For real frequencies, both modes are gapless and sound-like at low $k$, and are characterized by the density and the spin speeds of sound. The full spectrum of~(\ref{EqL}) has been solved in several references \cite{Kravchenko2009,Law2001,Takeuchi2010,Ishino2011}, with different scopes, and the general expression is cumbersome. 

\begin{figure}
  	\epsfig{file=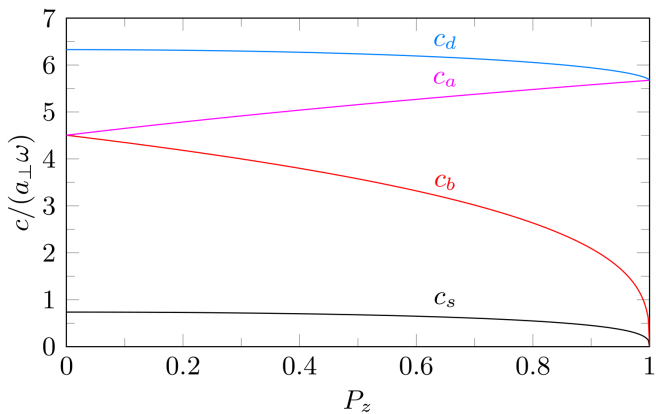,width=\linewidth,clip=true}
	\caption{Spin, density, and single-component speeds of sound. For concreteness, the densities have been calculated in Thomas-Fermi approximation (see Appendix).}\label{FigSpeeds}
\end{figure}

Let us review here two physical situations where the frequencies acquire a simple analytical form (the general solutions will be discussed in Secs.~\ref{SecPartStab} and \ref{SecDynInst}). The first case corresponds to $\mathbf{v_\sigma}=0$, that is the binary mixture at rest, and the dispersion relation takes the well-known form~\cite{Pethick}
\begin{equation}
\hbar\omega_{d(s)}=\sqrt{\frac{\hbar^2k^2}{2m} \left(\frac{\hbar^2k^2}{2m}+2mc_{d(s)}^2\right)} \label{EqDispRel}
\end{equation}
where the density ($d$) and spin ($s$) speeds of sound are given by
\begin{equation}
  c_{d(s)}^2=\frac{g_an_a+g_bn_b\pm\sqrt{(g_an_a-g_bn_b)^2+4n_an_bg_{ab}^2} }{2m}\label{EqSSS}
\end{equation}
where $n_\sigma=|\Psi_\sigma|^2$ are the equilibrium densities of the two components $\sigma=a,b$, and the $+$ and $-$ signs correspond to $c_d$ and $c_s$, respectively.
From Eq.~(\ref{EqDispRel}) one sees that, as already mentioned above, the excitation frequencies of both modes assume a linear dispersion $\omega_{d(s)}=c_{d(s)}k$ at low quasimomentum $k$. For repulsive interactions, which is the case under consideration, we have $c_d\ge c_s$. Figure~\ref{FigSpeeds} shows the behavior of $c_{d}$ and $c_{s}$ as a function of $P_z$. 
For completeness the single-component speeds of sound, $c_\sigma=\sqrt{g_\sigma n_\sigma/m}$, for $\sigma=a,b$, are also shown. 
To plot these velocities, the densities entering Eq.~(\ref{EqSSS}) have been calculated using a Thomas-Fermi approximation (see Appendix).
In the limit of $P_z\to1$ the density mode is dominated by the majority component and $c_a\to c_d$, while the spin mode is dominated by the minority component and $c_b\to c_s$.
Notice also from Eq.~(\ref{EqSSS}) that at the demixing transition point, i.e. $g_{ab}=g_{ab}^c$, the spin speed of sound vanishes for any polarization $P_z$, or equivalently the susceptibility of the mixture diverges.  Stability of persistent currents in this critical regime has been addressed in Refs.~\cite{Smyrnakis2009, Anoshkin2013} for a one-dimensional ring, and in Ref.~\cite{Bargi2010} in two dimensions.

The second case is for $\mb{v_{a}}=\mb{v_{b}}=\mb{v}$. It is easy to see this gives rise to a shift in the frequencies by the quantity $\mathbf{v}\cdot\mathbf{k}$, which has the role of a classical Doppler shift. An example of the behavior of the dispersion relations in this case is shown in Fig.~\ref{FigSpectrum}, calculated for Thomas-Fermi density profiles (see the Appendix). It can be seen that since the density mode is higher in energy, the effect of a nonzero velocity is small for our close-to-critical situation. In contrast, the dispersion relation of the spin mode is much more sensitive, and adding a nonzero velocity has strong consequences. In particular, for a large enough velocity the energy of the excitation can become negative, leading to an energetic instability, which we will show in Sec.~\ref{SecMisc} that is responsible to a great extent for the decay of the persistent currents.
Notice that in Fig.~\ref{FigSpectrum}, for convenience, we show the spectrum for velocity values $|\mb{v}|=v=2\pi\kappa\hbar/m$ corresponding to the quanta of circulation, $\kappa$, one would have in a ring geometry. % and for parameter values consistent with the numerical results shown below.

\begin{figure}
  	\epsfig{file=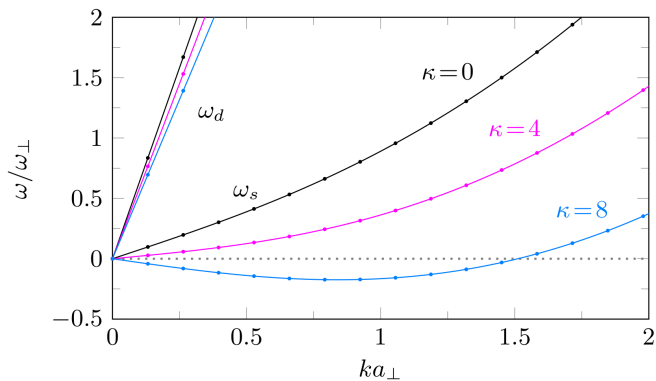,width=\linewidth,clip=true}
	\caption{Bogoliubov excitation spectrum for density ($\omega_d$) and spin ($\omega_s$) modes. The symbols correspond to the discretized values of $k$ (see text) arising from the ring geometry. For concreteness, the densities have been calculated in Thomas-Fermi approximation (see Appendix).
}\label{FigSpectrum}
\end{figure}

\subsection{Corrections to the speed of sound due to confinement}\label{SecCorrSound}

In the last paragraph we derived the speeds of sound for a uniform medium. When the system is confined the excitations still have a sound-like character provided the width of the cloud is large enough in the propagation direction and small enough in the transverse directions~\cite{Andrews1997,Dalfovo1999}. In this section we discuss the corrections to the frequencies Eq.~(\ref{EqDispRel}) that arise from a 2D ring geometry.

The first correction comes from the discretization of quasimomentum due to the multiply connected geometry, according to $k=\ell/R_0$, with $\ell$ the quantization number of the quasi-momentum ($k=\ell/R_0$) and $R_{0}$ the radius of the ring. To exemplify this correction, the discrete values of $k$ accessible to the system are represented as symbols in Fig.~\ref{FigSpectrum}. The effect of this correction on the stability criterion for persistent currents has been studied in Refs.~\cite{Smyrnakis2009, Anoshkin2013}. For the configuration that is addressed in this work this correction is negligible unless the polarization is very large. Indeed, we have checked that the correction is less than $1\%$ for $P_{z}<0.9$ and less than $10\%$ for $P_{z}<0.99$.

The second and most important correction to the speeds of sound Eq.~(\ref{EqSSS}) is brought about by the transverse (radial) degrees of freedom. For a single component in a three-dimensional harmonic trap the renormalizing factor ranges from unity in the non-interacting limit to $1/\sqrt{2}$ in the Thomas-Fermi limit~\cite{Zaremba1998,Kavoulakis1998,Stringari1998,Fedichev2001}. This correction depends only on the geometry and the density structure, not on the nature of the mode (spin or density) or the number of components (provided the mixture is miscible).  A ring trap (with $2\pi R_{0}$ larger than the radial width) can be thought of as a very long prolate trap with periodic boundary conditions, with the periodicity entering only as the discretization of $k$ discussed above. On the other hand, the density in the Thomas-Fermi limit for two components takes the same inverted parabola structure as for one component (see Appendix). Therefore, without loss of generality we calculate the Thomas-Fermi correction factor to the sound velocity for a single-component condensate confined in a two-dimensional prolate trap and apply it to the two-component case.

%We have considered a harmonically trapped 2d condensate (single component) in the Thomas-Fermi approximation, which captures the main features coming from the transversal degrees of freedom. 
%To calculate the correction to the sound velocity due to the transversal degrees of freedom in a toroidal trap we take into account the following considerations:
%\begin{enumerate}
% \item[(i)] The scaling factor will be at zeroth order the same as in a harmonic trap. That is, the bending of the condensate will only introduce a small correction to the factor obtained from a harmonic trap.
% \item[(ii)] 
%\end{enumerate}

To derive the correction factor of the speed of sound we follow Ref.~\cite{Stringari1998} and write the hydrodynamic equations for the density and the velocity in a two-dimensional system
\begin{align}
& \frac{\partial n}{\partial t} + \nabla_\perp(\mathbf{v}n)=0 \\
& m\frac{\partial \mathbf{v}}{\partial t} + \nabla_\perp\left[ V_{h}+gn-\frac{\hbar^2}{2m\sqrt{n}}\nabla_\perp^2\sqrt{n}+\frac{1}{2}mv^2 \right]=0
\end{align}
where $V_h=\frac{1}{2}m\omega_x^2x^2+\frac{1}{2}m\omega_y^2y^2$. We have in mind a situation satisfying $\omega_y\ll\omega_x$, which is the relevant one for a toroidal trap, where the $y$ coordinate corresponds to the azimuthal angle around the trap axis. Neglecting the quantum pressure term ($\sim\nabla_\perp^2\sqrt{n}$) the ground state at rest is characterized by an inverted parabola profile that extends between $Y^2=\pm\frac{2\mu}{m\omega_y^2}$ (and analogously along the $x$ axis). The system forms an ellipse on the $xy$ plane with minor and major axis respectively given by $X$ and $Y$. The chemical potential is given by $\mu=\hbar\omega_y\sqrt{N\tilde{g}/(\pi\lambda_\perp)}$, where $\tilde{g}=g/(\hbar\omega_y a_y^2)$ with $g$ the 2D coupling constant defined above and $a_y=\sqrt{\hbar/(m\omega_y)}$, and where we have introduced $\lambda_{\perp}=\omega_y/\omega_x$. 
% $\mu=\hbar\omega_y\sqrt{4N a/\lambda_{\perp}}$

By linearizing the hydrodynamic equations and combining them one finds the eigenvalue equation
\begin{equation}
 \omega^2\delta n = -\partial_x\left[ \frac{\mu-V_h}{m}\partial_x\delta n \right] -\partial_y\left[ \frac{\mu-V_h}{m}\partial_y\delta n \right]
\end{equation}
where $\delta n=n-n_0$, with $n_0$ the unperturbed density. Notice that we have assumed a temporal dependence of the perturbations  $\delta n\sim e^{i\omega t/\hbar}$ (analogously for the perturbation of the velocity).
We are interested in finding the lowest energy excitations, which consistently with our assumption $\omega_y\ll\omega_x$ will mainly come from the $y$-dependent part of the eigenvalue equation above. Neglecting thus all dependence of $\delta n$ on $x$, that is $\delta n=\delta n_0(y)$, and integrating the whole equation with respect to $x$ we find
\begin{equation}
 \omega\delta n_0=-\frac{1}{3}\omega_y^2(Y^2-y^2)\partial^2_y\delta n_0 + 2\omega_y^2 y\partial_y \delta n_0
\end{equation}
For the excitations localized at the center ($y\approx 0$) this yields
\begin{equation}
 \omega^2=\frac{1}{3}\omega_y^2Y^2k^2=\frac{2}{3}\frac{\mu}{m}k^2
\end{equation}
Using the well-known result $c_0=\sqrt{\mu/m}$ for the sound velocity in a uniform Bose-Einstein condensate we recover the dispersion relation $\omega=ck$ with 
\begin{equation}
  c=\sqrt{\frac{2}{3}}c_0\ .
\end{equation}
The correction factor to Eqs.~(\ref{EqSSS}) due to radial confinement is thus $\sqrt{2/3}$.
Notice that there might exist a small correction due to the anharmonicity of the trapping potential considered in the numerical results, Eq.~(\ref{EqPot}), and of the bending introduced by the ring geometry. These corrections, however, do not seem to have any appreciable effect in the simulations presented below. The main deviation from this factor would come from a density profile that was not close enough to the Thomas-Fermi limit (see for instance the results of \cite{Zaremba1998} in 3D).

\section{Persistent currents in the miscible regime}\label{SecMisc}

%\subsection{Spin speed of sound and stability}\label{SecMiscSSS}

We have discussed in Sec.~\ref{SecTheory} that Bogoliubov modes show that if the superfluid flows at a finite velocity the dispersion relation bends due to the Doppler shift (see Fig.~\ref{FigSpectrum}).
Consequently, when the flow velocity equals the stationary speed of sound, the dispersion relation touches the axis $\omega=0$, triggering an energetic Landau instability \cite{comment3}.
Since this instability appears first in the spin channel, it leads to the following criterion for the stability of persistent currents in mixtures: when the flow velocity is larger than the spin speed of sound, the currents become (energetically) unstable and thus decay. In this section we explore this criterion numerically.

\subsection{Stability diagram} \label{SecNum}

In order to discuss the stability diagram of the persistent currents and test the above criterion we have solved the GP equations with a vortex-like ansatz for the initial wave function~\cite{Abad2010}, 
\begin{equation}
 \Psi_{\sigma}(\tau=0)= \psi_{\sigma} \left(\frac{x+iy}{\sqrt{x^2+y^2}}\right)^{\kappa_{\sigma}}\label{EqPsi}
\end{equation}
with $\sigma=a,b$, $\tau$ the imaginary time variable, and $\kappa_{a}=\kappa_{b}=\kappa$. The system is then allowed to evolve freely in imaginary time until convergence is reached. 
For all simulations the initial trial wave functions, $\psi_{\sigma}$, have been built from both random density and phase distributions, in order to prevent the algorithm from reaching false metastable states. We have checked that the virial theorem for the trapping potential~(\ref{EqPot}) is always fulfilled when convergence is reached, namely
\begin{equation}
	\sum_{\sigma=a,b}\left(2E_{\text{kin},\sigma} - 2E_{\text{trap},\sigma} + 2E_{\text{int},\sigma} + \delta E_\sigma \right) + 2E_{\text{int},ab}=0
\end{equation} 
where $E_{\text{kin}}$, $E_{\text{trap}}$ and $E_{\text{int}}$ are, respectively, the kinetic, trapping and non-linear interaction energy terms, and $\delta E$ comes from the anharmonicity of potential~(\ref{EqPot}) and is given by
\be
  \delta E_\sigma= 2 \int d{\bf r_\perp}|\Psi|^2 V_0\left(1+\frac{2r_\perp^2}{\sigma_0^2}\right)e^{-2r_\perp^2/\sigma_0^2}
\ee
After convergence, we calculate the expectation value of the angular momentum per particle, $L_z^{(\sigma)}=\braket{\Psi_\sigma|-i\hbar\partial_\varphi|\Psi_\sigma}/N_\sigma$, and the circulation integral, $\Gamma_\sigma=\oint \mathbf{v}_\sigma\cdot d\ell$, with $\mathbf{v}_\sigma$ the velocity field and with the integral evaluated in a closed circuit around the central hole. 

\begin{figure}
	\epsfig{file=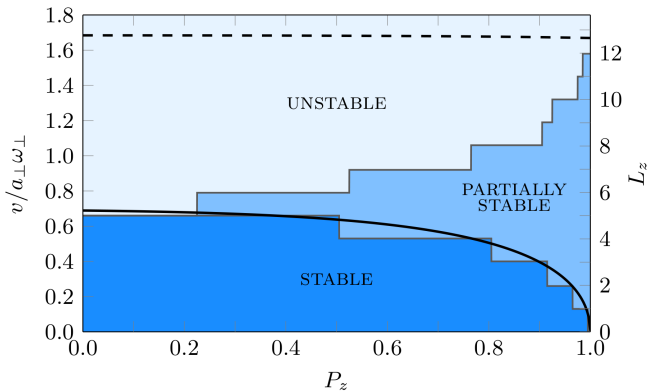, width=\linewidth, clip=true}
	\caption{Stability diagram in the miscible regime as obtained from imaginary-time simulations. The solid line represents the spin speed of sound $c_{s}$ [see Eq.~\eqref{EqSSS}] computed at the density maxima by taking into account the corrective factor for 2D geometries. The upper dashed line represents the boundary for dynamical instability when $v_{b}=0$ as obtained from Bogoliubov analysis.}\label{Figvelocity}
\end{figure}

The results are shown in Fig.~\ref{Figvelocity} as a function of $P_z$. The left $y$-axis shows the initial velocity and the right $y$-axis the initial angular momentum per particle, which is quantized in multiples of $\kappa$. The metastability of the intial states is shown as different shaded regions, corresponding to different stabilty regimes.
The supercurrent is stable (dark region) if the velocity at the density maximum is smaller than the spin sound velocity (black solid line), in good agreement with the above criterion relating supercurrent instability to the Landau instability of the spin mode (see also \cite{Anoshkin2013}). The spin speed of sound has been renormalized by the factor $\sqrt{2/3}$ that takes into account the effect of the transverse width of the condensate (Sec.~\ref{SecCorrSound}). 
%which has been approximated by a local density profile. This reduces the spin speed of sound by a factor $\sqrt{2/3}$ (see Supplemental Material), which is analogous to the factor $1/\sqrt{2}$ found in the three-dimensional case~\cite{Zaremba1998,Stringari1998,Fedichev2001}.

Furthermore, the numerical simulations allow us to distinguish between two unstable regions: (i) a {\it partially stable} one, where the current in the majority component is stable, while it decays in the minority component; (ii) an {\it unstable} one where both currents are unstable. While the boundary between the partially stable and the stable region is determined by the spin speed of sound, the boundary between the partially stable and the unstable region is not fixed by any universal criterion and its exact position depends on the geometry of the system, as is discussed in the next paragraph. 

\subsection{Partially stable region}\label{SecPartStab}

The presence of the partially stable region is in agreement with the fact that the spin mode is dominated by the minority component in the limit of $P_z\to1$ (see also Fig.~\ref{FigSpeeds}). In physical terms, the minority component is more sensitive to spin excitations, whereas the majority component becomes more stable, being less affected by spin-density excitations.
In mathematical terms, further insight is provided by considering the case where the two components have different velocities, $v_a\neq v_{b}$ (notice that the velocities have nonzero components mainly in the azimuthal direction, since they show a vortex structure). It is easy to prove that once the minority component has lost a part of its initial angular momentum, the dispersion relation is no longer given by Eq.~(\ref{EqSSS}) and a non-linear Doppler shift is originated by the velocity difference. As a result, the system becomes more stable. 

\begin{figure}
	\epsfig{file=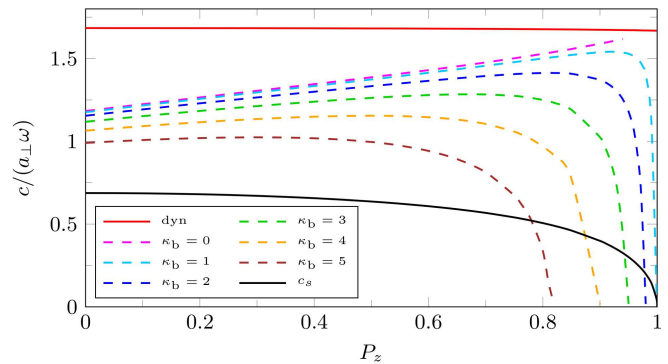,width=\linewidth,clip=true}
	\caption{Lines of energetical instability for $v_{b}=\hbar\kappa_b/mR_{0}$  for different $\kappa_b$. The instability corresponds to the maximum velocity for the $a$ component that leads to a positive spin-mode frequency.}\label{vadiffvb}
\end{figure}

An example of this analysis is shown in Fig.~\ref{vadiffvb}. This figure shows the maximum velocity that component $a$ can carry for a fixed (quantized) initial velocity of $b$, such that the energy of the spin excitations, $\omega$ in Eq.~(\ref{EqBog}),  remains positive. 
%An example of what happens when the velocities are different is shown in Fig.~\ref{vadiffvb}. 
The different curves have been obtained by diagonalizing the operator $\mathcal{L}$, see Eq.~(\ref{EqL}), with the parameters and the densities taken from the ground state of the GP equations in the absence of currents. Also, the factor $\sqrt{2/3}$ (see Sec.~\ref{SecCorrSound}) has been applied to all curves. For comparison, we have also plotted the spin sound velocity for equal flow velocities, $c_{s}$, and the line of dynamical instability for $v_{b}=0$ (see Sec.~\ref{SecDynInst}). We see from the figure that, at fixed $P_{z}$, as the velocity difference grows (that is, $\kappa_{b}$ decreases), the allowed maximum velocity for component $a$ is larger. This means that the superflow can be stabilized by losing velocity in one of the components (in this case the minority component $b$).
This argument justifies the presence of the partially stable region in Fig.~\ref{Figvelocity}.

\begin{figure*}
  \epsfig{file=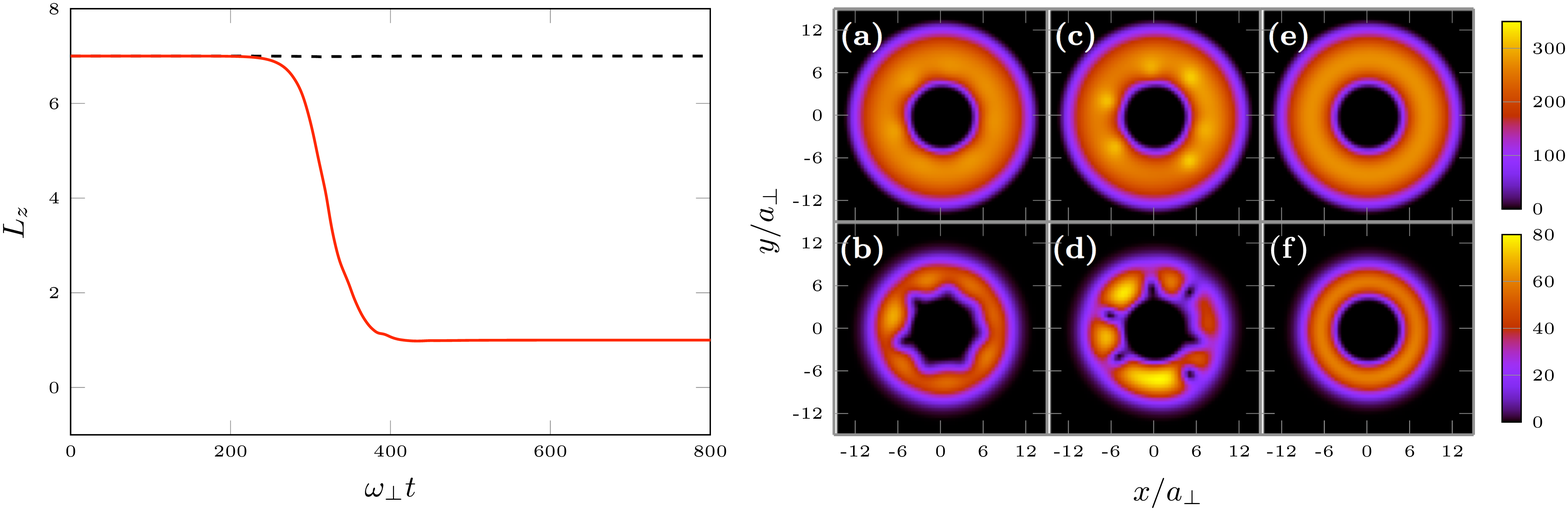,width=\textwidth,clip=true}
  \caption{Left panel: time-dependence of angular momentum of components $a$ (dashed line) and $b$ (solid line) in real-time dynamics in the partially stable region. Right panels: density snapshots of components $a$ (upper row) and $b$ (lower row) at times: $t=260\omega_\perp^{-1}$ in panels (a) and (b), $t=305\omega_\perp^{-1}$ in panels (c) and (d), and $t=560\omega_\perp^{-1}$ in panels (e) and (f). For this case $P_z=0.8$ and $\kappa=7$.}\label{FigPsDyn}
\end{figure*}

To test it further and to avoid any spurious effects of imaginary time, we have run real-time simulations in the partially stable region, adding a dissipation term in Eqs.~(\ref{tdgpa})-(\ref{tdgpb}) in analogy to what has been done in Ref.~\cite{Yakimenko}. We have also added a very small random noise to the potential to make the energetic instability appear in a shorter time scale (we have checked that the same result is obtained without this random noise). The wave functions have been renormalized at every time step to the initial number of particles, following what was done in Ref.~\cite{Tsubota}, and the dissipation parameter has been taken to be $\gamma=0.08$, as in \cite{Yakimenko}. We show the results in Fig.~\ref{FigPsDyn}. In the left panel the time evolution of the angular momentum of components $a$ and $b$ is  shown. While the currents in component $a$ remain stable throughout all the dynamics, the currents in component $b$ decay. This decay is indeed induced by the crossing of vortices across the ring, as can be seen in the density snapshots of Fig.~\ref{FigPsDyn}, for the majority (upper row) and the minority (lower row) components. 
These snapshots show the dynamical process explained above very clearly: first the spin instability kicks in as out-of-phase density oscillations in the azimuthal direction, as seen in panels (a) and (b); since the minority component is more sensitive to this perturbation, its density oscillations grow enough to allow the penetration of vortices inside the ring, panels (c) and (d); finally, after losing angular momentum, the system is stabilized through the new stability criterion shown in Fig.~\ref{vadiffvb}, panels (e) and (f). %Notice that the vortices escape preserving the symmetry of Eqs.~(\ref{tdgpa})-(\ref{tdgpb}), since we have not added any external noise and the instability is of the energetic type, that is it does not grow exponentially from a seed that can be numerical noise (as happens with the dynamical instability of Sec.~\ref{SecDynInst}).

%It is clearly seen in this simulation that when the spin instability kicks in it is the minority component (being more sensitive to it) that loses the circulation through the emission of quantized vortices that cross the ring. Then, the stability criterion is changed and the currents can be either stable (thus the original currents are partially stable) or unstable, in which case also the majority component loses its angular momentum. We have checked that the stability criterions for $v_{a}$ shown in Fig.~\ref{vadiffvb} are consistent with imaginary time simulations when the intial wave function is given by Eq.~(\ref{EqPsi}) with $\kappa_{a}\neq\kappa_{b}$.

Before ending our discussion of the partially stable region, let us comment on the one-dimensional (1D) limit. In this case the solutions of Eq.~(\ref{EqBog}) are exact, in the sense that there is no renormalization factor accounting for the external degrees of freedom. The effect of going to the 1D-limit is to increase the width of the partially stable region. This can be seen in Table~\ref{Table}, where we compare the width obtained from numerical simulations of the 1D GP equations with that of the 2D GP equations (in units of the corresponding sound velocities at $P_z=0$). 
Therefore, the partially stable region is not a feature of an extended geometry, but it is present also in one-dimensional systems. The universality of the 1D limit makes this result relevant to  coupled Luttinger liquids (see, e.g., \cite{Yu1992, Viefers2000}).

\begin{table}[h]\caption{Width of the partially stable region}\label{Table}
\begin{tabular}{c|cc}
 $P_z$ & \hspace{1em}$\Delta^{1D}/c_s^{1D}$ & \hspace{1em}$\Delta/c_s$ \\\hline
% $0.3$ & \hspace{1em}$0.74$ & \hspace{1em}$0.42$ \\
% $0.6$ & \hspace{1em}$1.54$ & \hspace{1em}$1.06$ \\
% $0.9$ & \hspace{1em}$4.00$ & \hspace{1em}$2.25$ 
  $0.10$ & $0.2398$ & $0.0000$ \\
  $0.20$ & $0.2398$ & $0.0000$ \\
  $0.30$ & $0.4797$ & $0.1911$ \\
  $0.60$ & $0.9594$ & $0.5734$ \\
  $0.90$ & $1.4391$ & $0.9557$ \\
  $0.95$ & $1.6789$ & $1.5291$
\end{tabular}
\end{table}

\subsection{Dynamical instability}\label{SecDynInst}

\begin{figure}
	\epsfig{file=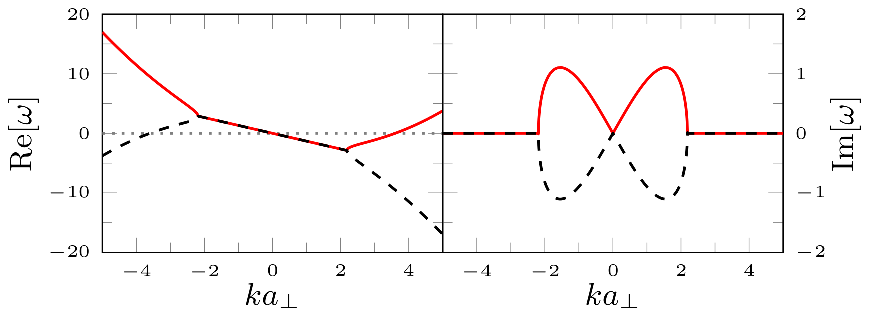, width=\linewidth, clip=true}
	\epsfig{file=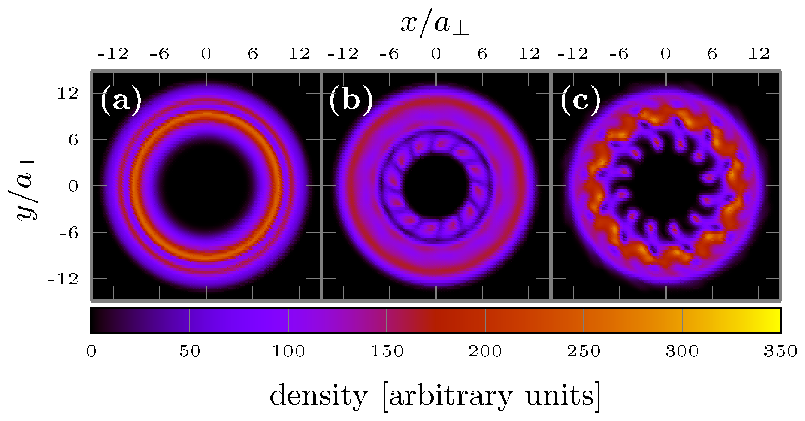, width=\linewidth, clip=true}
	\caption{Dynamical instability for $\kappa_a=20$, $\kappa_b=0$. Top row: Real and imaginary parts of the dispersion relation for the spin mode. When the imaginary part vanishes positive- (negative-) norm solutions are represented in solid (dashed).
Bottom rows: Snapshots of majority component densities during real-time dynamics, at times $t=31.5$~ms (a), $t=47.7$~ms (b) and $t=57.4$~ms (c).}\label{FigDynInst}
\end{figure}

The energetic instability discussed above, although being the relevant one when the two superfluids have the same velocity, is not the only mechanism that can trigger decay of persistent currents in a binary mixture.
Indeed, when $|{v}_a-{v}_b|$ exceeds some threshold the eigenfrequencies corresponding to the spin-density mode acquire an imaginary part, leading to an exponential growth of the spin excitations that makes the flow dynamically unstable. 
The existence of a dynamical instability for different flow velocities is a more general result and it is due to the breaking of Galilean invariance. This has been recently discussed in spin-orbit coupled condensates~\cite{sandroso}.
In the context of binary mixtures, this instability is known as counterflow instability, and has been addressed both experimentally~\cite{Hammer2011,Hoefer2011} and theoretically~\cite{Law2001,Kravchenko2009,Takeuchi2010,Ishino2011}. The structure of the complex eigenfrequencies is illustrated in the top panels of Fig.~\ref{FigDynInst}: the real part (left panel) is nonzero in the limit of small $k$, in contrast to the case of the demixing instability driven by inter-species interaction.

%In one-dimensional condensates, the counterflow instability has been observed through the creation of a dark-bright soliton train \cite{Hammer2011}. 
To better characterize how the dynamical instability appears in a toroidal trap, we have performed real-time simulations of the Eqs.~(\ref{tdgpa})-(\ref{tdgpb}), imposing initial winding numbers $\kappa_a=20$, $\kappa_b=0$, which correspond to a velocity $v_{a}$ much larger than the critical velocity (Fig.~\ref{Figvelocity}, dashed line).
The initial state consists on the converged solutions of the GP equations describing the mixture at rest, on which we have added an initial vortex-like phase following Eq.~(\ref{EqPsi}). 

Selected snapshots of the majority component density are shown in Fig.~\ref{FigDynInst}, showing three different regimes: a first stage with radial breathing (as recently discussed in \cite{Murray2013} for a one component BEC), a second stage where the spin instability kicks in and deforms the condensates, and a third stage in which vortices enter the BECs and stabilize the angular momentum at $L_z^{(a)}=L_z^{(b)}=L_z/2$.
The maxima in the density of one component coincide with the minima in the other, thus confirming that the instability is driven by the spin-density mode. 
Notice that in absence of dissipation the total angular momentum is conserved; however, adding a small imaginary term in the left-hand-side of Eqs.~(\ref{tdgpa})--(\ref{tdgpb}) we obtain dissipative dynamics where both energy and angular momentum decrease in time, and vortices are then able to fully cross the torus (after a certain time).

\section{Persistent currents in the immiscible regime}\label{SecPhSep}
\begin{figure}[h]
	\epsfig{file=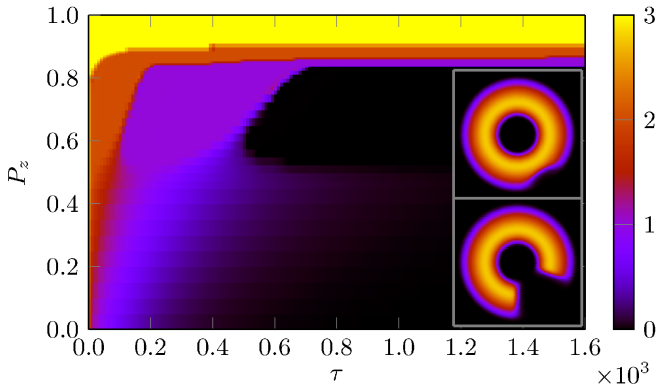, width=\linewidth, clip=true}
\caption{Stability diagram in the phase separated regime for the majority component, for different values of the polarization and as a function of the imaginary time, $\tau$. The color scale represents the value of the circulation. In the insets, we have plotted the final density of the majority component for $P_z=0.95$ (top) and $P_z=0.6$ (bottom).}\label{Figcurrent}
\end{figure}

The physics of persistent current decay is very different in the phase-separated regime. Since here simple analytical arguments cannot be used and the results strongly depend on the geometry of the system, we only rely on numerics. The parameters are the same as in the miscible case, but $g_{ab}/g=1.02$. The insets of Fig.~\ref{Figcurrent} show the density of the majority component for two different polarizations. The minority component (not shown) occupies the empty regions of the ring. When $P_z$ is small, the two components occupy two sectors of the ring whose length is determined by the value of $P_{z}$ (bottom inset). Conversely, when $P_{z}$ is large, the minority component occupies only a small region, creating a barrier (or weak link, depending on the value of the penetration length) for the majority component (top inset). These two kinds of density distributions correspond, respectively, to ground states characterized by $\mu_a=\mu_b$ and by $\mu_a>\mu_b$.

In the phase separated regime the current in the minority component is always unstable, since it occupies a simply connected region in the torus. The persistence of the current in the majority component depends instead on satisfying two conditions: (i) it occupies a multiply connected region, and (ii) the barrier formed by the minority component is small enough to prevent vortices from escaping (similarly to what happens in weak-link systems \cite{Piazza2009,Ramanathan2011,Wright2013} and in dipolar condensates~\cite{Abad2010}). 
Such conditions are fulfilled only for high polarizations, as illustrated in Fig.~\ref{Figcurrent}, where the circulation of the majority component is shown in color scale as a function of $P_{z}$ and of the imaginary time. 
Despite not being real time, imaginary time evolution gives an idea of whether the system reaches its ground state, or else remains blocked in a metastable state. Qualitatively similar behavior is expected in an experiment due to dissipative effects. In the present situation, we identify three regimes: For $P_z > 0.9$ persistent currents are stable, for $0.6<P_z<0.9$ decay consists in a series of jumps through states with integer circulation, and for $P_z < 0.6$ decay is continuous.

The situation is different when the intra-species coupling constants $g_a$ and $g_b$ are different. In this case the ground state consists of two concentric rings similarly to what is found in purely harmonic traps (see for instance \cite{Svidzinsky}). In this configuration our simulations show that the persistent flow is maintained to a higher degree. However, a systematic numerical analysis is complicated by the formation of domain walls that produce quasi-degenerate configurations whose energy difference is smaller than the numerical precision.

\section{Persistent currents for coherently coupled BECs}\label{SecCoherent}

For the sake of completeness, we briefly discuss the case in which particle exchange is allowed by the presence of a linear coupling $\Omega\Psi_{b}$ ($\Omega\Psi_{a}$) in Eq.~\eqref{tdgpa} (Eq.~\eqref{tdgpb}). The miscible-immiscible transition is replaced by a second order one, which  takes place at $g_{ab}^{c}=g+2\Omega/n$, and which separates a neutral from a polarized regime (see~\cite{Abad2013} and references therein). A gap opens in the spin-density mode and the Landau critical velocity becomes larger than in the binary mixture. Consequently, the spinor two-component condensate becomes stable in configurations where the mixture was unstable. Stability is reinforced by the presence of $\Omega$, which ensures $P_z=0$. This has been numerically checked for values of the parameters corresponding to a miscible mixture ($g_{ab}<g$) as well as to an immiscible one ($g<g_{ab}<g+2\Omega/n$).
This result fully agrees with the experimental observations of \cite{Beattie2013}. In the polarized regime, the criterion for stability is more complex since the neutral and polarized phases always coexist in the trap \cite{Abad2013}. However, phase coherence still guarantees that the two species decay together.

\section{Conclusions.}\label{SecConclusions}

We have studied the stability of persistent currents in two-component condensates, both in the miscible and in the immiscible regimes. In general terms, in the miscible regime persistent currents decay when the flow velocity is larger than the spin speed of sound (see also \cite{Smyrnakis2009,Bargi2010,Anoshkin2013}). 
At a first glance, this criterion seems to predict a behavior opposite to the experimental results~\cite{Beattie2013}. 
However, as we have argued in this article, a more careful analysis shows that a
new region in the stability diagram is present, which would point in the right direction to explain the results of~\cite{Beattie2013}.
Indeed, by analyzing the metastable states of the system and performing a linear stability analysis on them, we have found that for high enough polarizations there exists a partially stable region where the superflow in the minority component decays while in the majority component it remains stable. 
This is compatible with the experimental results in \cite{Beattie2013}, where the population in the minority component for high polarizations could not be determined in a precise way: the system could end up in the partially stable regime but be detected as fully stable. 
This mechanism is physically justified by the fact that the spin-density mode affects more strongly the minority component than the majority component, and we have seen that it is indeed what happens in a real-time dynamics in the presence of dissipation. 
In an indirect way, this mechanism was seen in the numerical work of \cite{Yakimenko}, where the majority component stabilizes at a nonzero value of the angular momentum, while the minority component loses it completely. 
The geometry of the trap can change to a great degree the shape of the partially stable region, making it steeper for high polarizations, as seen from Table~\ref{Table}. Therefore, to explore the accuracy of our predictions and understand the full physical picture, more experiments should be carried out, especially for different initial circulations. 

Let us remark here that while in our semiclassical treatment persistent currents remain stable for an infinite time in the absence of dissipation, quantum and thermal fluctuations can drive their decays. However, while the dynamics of the decay toward the ground state will be affected by those fluctuations (see for instance the experiment \cite{Moulder2012}), the stability properties of the states will not be significantly modified. Thus the main features of the results presented in this article will be recovered in a more sophisticated analysis taking into account also those effects.

In this article, we have also discussed the dynamical instability arising from a large velocity difference between the two components. On the other hand, we have analyzed the stability criterion of persistent currents in the immiscible regime, where (for the equal mass and equal interspecies interactions) the results depend strongly on the density structure. For high polarizations, the minority component acts as a small barrier and the majority component can stabilize the currents.

Finally, we have discussed that the presence of a coherent coupling stabilizes persistent currents, since a gap in the spin channel opens and the energetic instability appears at higher values of the flow velocity. This results are in agreement with the observations in \cite{Beattie2013} when the coherent coupling was kept on during the experiment.

%Our work can help in the understanding of the experimental results reported in Ref.~\cite{Beattie2013}, which do not agree with existing theoretical results~\cite{Smyrnakis2009,Bargi2010,Anoshkin2013}.
%However in \cite{Beattie2013} a single flow velocity ($\kappa=3$) is considered; thus, in order to test our prediction, further experimental investigations are required. 

\acknowledgments
We acknowledge stimulating discussions with I. Carusotto,  Z. Hadzibabic and T. Ozawa. This work has been financially supported by ERC through the QGBE grant and by Provincia Autonoma di Trento.

\appendix

\section{Thomas-Fermi approximation in a ring trap}

\begin{figure}
 \epsfig{file=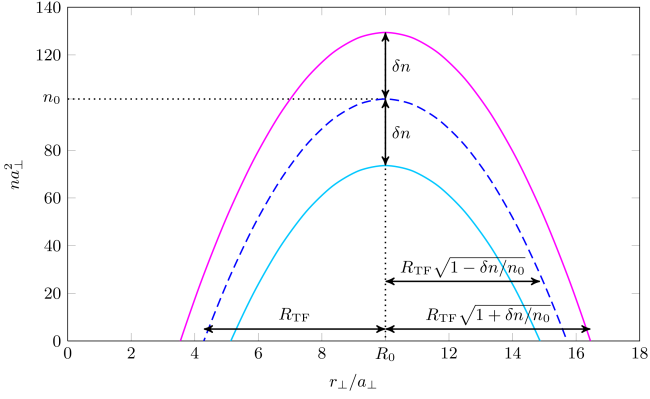,width=\linewidth,clip=true}
 \caption{Thomas-Fermi density profiles for a two-component mixture in a displaced harmonic trap. Dashed line: $N_a=N_b$. Solid lines: $P_z=0.4$.}\label{SuppFigTF}
\end{figure}

In this Appendix we characterize the Thomas-Fermi (or local density approximation) solution of a two-component BEC in a ring trap. This provides with simple analytical results that can be used to obtain an approximation to various quantities.
To this aim, it is convenient to assume a confining potential that is a displaced harmonic trap, as has been done in \cite{Bargi2010},
\begin{equation}
 V(r_\perp)=\frac{1}{2}m\omega_\perp^2\,(r_\perp-R_0)^2
\end{equation}
where $m$ is the atomic mass, $\omega_\perp$ is the trapping frequency, $r_\perp=\sqrt{x^2+y^2}$ is the radial coordinate, and $R_0$ is the position of the potential minimum. Neglecting the quantum pressure term in the Gross-Pitaevskii equations (local density approximation), and assuming equal intra-species scattering lengths ($g_a=g_b=g$), the densities for components $\sigma=a,b$ are given by
\begin{equation} 
n_\sigma=n_0\left[1-\left(\frac{r_\perp-R_0}{R_{TF}}\right)^2\right]\pm \delta n\,, 
\end{equation}
with $\pm$ corresponding respectively to $a$ and $b$. In the last expression we have used the central density, the Thomas-Fermi radius and the density difference, given respectively by
\begin{align}
 & n_0= \frac{\mu_a+\mu_b}{g+g_{ab}}\\
 & R_{TF}^2 =\frac{\mu_a+\mu_b}{m\omega_\perp^2} \\
 & \delta n=\frac{\mu_a-\mu_b}{g+g_{ab}}
% n_0= \frac{\mu_0}{g+g_{ab}}\left[\left(\frac{2N_a}{N}\right)^{2/3}+ \left(\frac{2N_b}{N}\right)^{2/3}\right]
\end{align}
%the Thomas-Fermi radius 
%\begin{equation}
% R_{TF}^2 =\frac{\mu_a+\mu_b}{m\omega_\perp^2}
%\end{equation}
%and the density difference given by
%\begin{equation}
% \delta n=\frac{\mu_a-\mu_b}{g+g_{ab}}
%\end{equation}
The quantity $R_{TF}$ corresponds to the Thomas-Fermi radius of the total density. Since the two components can be differently populated, the radial extent of the corresponding clouds can also be different, which allows us to define inner ($+$) and outer ($-$) radii
\begin{align}
   &R_\pm^a=R_0\pm R_{TF}\sqrt{1+\frac{\delta n}{n_0}}\\
   &R_\pm^b=R_0\pm R_{TF}\sqrt{1-\frac{\delta n}{n_0}}
\end{align} 
Normalization of the densities to $N_a$ and $N_b$ gives chemical potentials
\begin{align}
 \mu_a&= \frac{\mu_0}{g+g_{ab}}\left[g\left(\frac{2N_a}{N}\right)^{2/3}+ g_{ab}\left(\frac{2N_b}{N}\right)^{2/3}\right] \\
 \mu_b&= \frac{\mu_0}{g+g_{ab}}\left[g_{ab}\left(\frac{2N_a}{N}\right)^{2/3}+ g\left(\frac{2N_b}{N}\right)^{2/3}\right] 
\end{align}
where $\mu_0$ is given by
\begin{equation}
%	\mu_0 = \frac{1}{2}\hbar\omega_\perp\left[ \frac{3}{4}\frac{a+a_{ab}}{R_0}N \right]^{2/3}
	\mu_0 = \frac{1}{2}\hbar\omega_\perp\left[ \frac{3}{16\pi}\frac{g+g_{ab}}{\hbar\omega_\perp a_\perp^2}\frac{a_\perp}{R_0}N \right]^{2/3}
\end{equation}
and corresponds to the chemical potential of the symmetric mixture ($N_a=N_b$).
An example of how this density profile looks like is given in Fig.~\ref{SuppFigTF}, where the relevant parameters are indicated. 

%\bibitem{Malet2010} F. Malet, G. M. Kavoulakis, and S. M. Reimann, Phys. Rev. A {\bf 81}, 013630 (2010).

%\begin{figure}[h!]
%	\epsfig{file=figures/soundVaVb/soundvavb.eps, width=\linewidth, clip=true}
%	\caption{%(Color online) Stability diagram predicted by the numerical simulations. The green shaded area represents the full stability region. The result of Eq.~\eqref{EqSSS} for the density maxima is depicted as a solid line. The region marked as {\it partially stable} indicates configurations where currents in the minority component ($b$) decay while in the majority ($a$) remain stable. As a guide to the eye we have drawn the line (dashed) defining its upper bound. The region marked {\it unstable} indicates the points where both components decay. The upper solid line represents dynamical instability for $\kappa_b=0$ as obtained from Bogoliubov analysis. 
%}\label{Figsoundvavb}
%\end{figure}


\thebibliography{99}

\bibitem{Bloch} F. Bloch, Phys. Rev. A {\bf 7}, 2187 (1973).

\bibitem{Moulder2012} S. Moulder, S. Beattie, R. P. Smith, N. Tammuz, and Z. Hadzibabic, Phys. Rev. A {\bf 86}, 013629 (2012).

\bibitem{Benakli} M. Benakli, S. Raghavan, A. Smerzi, S. Fantoni, and S. R. Shenoy, Europhys. Lett. {\bf 46}, 275 (1999).

\bibitem{Tempere} J. Tempere. J. T. Devreese, and E. R. I. Abraham, Phys. Rev. A {\bf 64}, 023603 (2001).

\bibitem{Dora} P. Capuzzi and D. M. Jezek, J. Phys. B: At. Mol. Opt. Phys. {\bf 42}, 145301 (2009).

\bibitem{Smyrnakis2009} J. Smyrnakis, S. Bargi, G. M. Kavoulakis, M. Magiropoulos, K. K\"arkk\"ainen, and S. M. Reimann, Phys. Rev. Lett. {\bf 103}, 100404 (2009).

\bibitem{Bargi2010} S. Bargi, F. Malet, G. M. Kavoulakis, and S. M. Reimann, Phys. Rev. A {\bf 82}, 043631 (2010).

\bibitem{Anoshkin2013} K. Anoshkin, Z. Wu, and E. Zaremba, Phys. Rev. A {\bf 88}, 013609 (2013).

\bibitem{Wu2013} Z. Wu and E. Zaremba, Phys. Rev. A {\bf 88}, 063640 (2013). 

\bibitem{Smyrnakis2014} J. Smyrnakis, M. Magiropoulos, N. K. Efremidis, and G. M. Kavoulakis, arXiv:1312.0869v2.

\bibitem{Beattie2013} S. Beattie, S.  Moulder, R. J. Fletcher, and Z. Hadzibabic, Phys. Rev. Lett. {\bf 110}, 025301 (2013).

\bibitem{Yakimenko} A. I. Yakimenko, K. O. Isaieva, S. I. Vilchinskii, and M. Weyrauch, Phys. Rev. A {\bf 88}, 051602(R) (2013).

\bibitem{Leggett} {\it Quantum Liquids. Bose Condensation and Cooper Pairing in Condensed-Matter Systems}, A. J. Leggett (Oxford University Press, 2006).

%\bibitem{foot_rot} {It is worth noticing that the physics underlying persistent currents is conceptually different from that expected in rotating systems~\cite{Bargi2007,Halkyard2010,Shimodaira2010}, since the former are excited states of the system while the latter correspond to the thermodynamical ground state in the rotating frame (see for instance~\cite{Leggett}).}
 
%\bibitem{Bargi2007} S. Bargi, J. Christensson, G. M. Kavoulakis, and S. M. Reimann, Phys. Rev. Lett. {\bf 98}, 130403 (2007).
 
%\bibitem{Halkyard2010} P. L. Halkyard, M. P. A. Jones, and S. A. Gardiner, Phys. Rev. A {\bf 81}, 061602(R) (2010).

\bibitem{Ramanathan2011}  A. Ramanathan, K. C. Wright, S. R. Muniz, M. Zelan, W. T. Hill III, C. J. Lobb, K. Helmerson, W. D. Phillips, and G. K. Campbell, Phys. Rev. Lett. {\bf 106}, 130401 (2011).

\bibitem{Wright2013} K. C. Wright, R. B. Blakestad, C. J. Lobb, W. D. Phillips, and G. K. Campbell, Phys. Rev. Lett. {\bf 110}, 025302 (2013).

\bibitem{Murray2013} N. Murray, M. Krygier, M. Edwards, K. C. Wright,  G. K. Campbell, and C. W. Clark, ArXiv:1309.2257.

\bibitem{Svidzinsky} A. A. Svidzinsky and S. T. Chui, Phys. Rev. A {\bf 67}, 053608 (2003). 

\bibitem{Shimodaira2010} T. Shimodaira, T. Kishimoto, and H. Saito, Phys. Rev. A {\bf 82}, 013647 (2010).

\bibitem{Mazets2002} I. E. Mazets, Phys. Rev. A {\bf 65} 033618 (2002).

\bibitem{Takeuchi2013} H. Takeuchi and K. Kasamatsu, Phys. Rev. A {\bf 88}, 043612 (2013).

\bibitem{Takeuchi2010} H. Takeuchi, S. Ishino, and M. Tsubota, Phys. Rev. Lett. {\bf 105}, 205301 (2010).

\bibitem{Ishino2011} S. Ishino, M. Tsubota, and H. Takeuchi, Phys. Rev. A {\bf 83}, 063602 (2011).

\bibitem{Law2001} C. K. Law, C. M. Chan, P. T. Leung, and M.-C. Chu, Phys. Rev. A {\bf 63}, 063612 (2001).

\bibitem{Kravchenko2009} L. Y. Kravchenko and D. V. Fil, J. Low Temp. Phys. {\bf 155}, 219 (2009).

\bibitem{Pethick} {\it Bose-Einstein Condensation in Dilute Gases}, C. J. Pethick and H. Smith (Cambridge University Press, 2nd Edition, 2008).

\bibitem{Andrews1997} M. R. Andrews, D. M. Kurn, H.-J. Miesner, D. S. Durfee, C. G. Townsend, S. Inouye, and W. Ketterle, Phys. Rev. Lett. {\bf 79}, 553 (1997).

\bibitem{Dalfovo1999} F. Dalfovo, S. Giorgini, L. Pitaevskii, and S. Stringari, Rev. Mod. Phys. {\bf 71}, 463 (1999).

\bibitem{Zaremba1998} E. Zaremba, Phys. Rev. A {\bf 57}, 518 (1998).

\bibitem{Kavoulakis1998} G. M. Kavoulakis and C. J. Pethick, Phys. Rev. A {\bf 58}, 1563 (1998).

\bibitem{Stringari1998} S. Stringari, Phys. Rev. A {\bf 58}, 2385 (1998).

\bibitem{Fedichev2001} P. O. Fedichev and G. V. Shlyapnikov, Phys. Rev. A {\bf 63}, 045601 (2001).

\bibitem{comment3} Notice that if one considers quantized $k$ values as discussed in Sec.~\ref{SecCorrSound} the instability criterion changes to the velocity values for which the frequency corresponding to a discrete $k$ value becomes negative. 

\bibitem{Abad2010} M. Abad, M. Guilleumas, R. Mayol, M. Pi, and D. M. Jezek, Phys. Rev. A {\bf 81}, 043619 (2010).

\bibitem{Tsubota} M. Tsubota, K. Kasamatsu, and M. Ueda, Phys. Rev. A {\bf 65}, 023603 (2002).

\bibitem{Yu1992} N. Yu and M. Fowler, Phys. Rev. B {\bf 45}, 11795 (1992).

\bibitem{Viefers2000} S. Viefers, P. Singha Deo, S. M. Reimann, M. Manninen, and M. Koskinen, Phys. Rev. B {\bf 62}, 10668 (2000).

\bibitem{sandroso} T. Ozawa, L. P. Pitaevskii, and S. Stringari, Phys. Rev. A {\bf 87}, 063610 (2013).

\bibitem{Hammer2011} C. Hammer, J. J. Chang, P. Engels, and M. A. Hoefer, Phys. Rev. Lett. {\bf 106}, 065302 (2011).

\bibitem{Hoefer2011} M. A. Hoefer, J. J. Chang, C. Hammer, and P. Engels, Phys. Rev. A {\bf 84}, 041605(R) (2011).

%\bibitem{Shevchenko2007} S. I. Shevchenko and D. V. Fil, Journal of Experimental and Theoretical Physics {\bf 105}, 135 (2007).

\bibitem{Piazza2009} F. Piazza, L. A. Collins, and A. Smerzi, Phys. Rev. A {\bf 80}, 021601(R) (2009).

\bibitem{Abad2013} M. Abad and A. Recati, Eur. Phys. J. D {\bf 67}, 148  (2013).
\end{document}